\documentclass{book}
\usepackage{soul}
\sethlcolor{white}
\usepackage{xcolor}
\usepackage{graphicx}
\usepackage{comment}
\usepackage{booktabs}
\usepackage{amsmath,amssymb,amsfonts}
\usepackage{caption}
\usepackage{subcaption}
\usepackage[a4paper, left=1in]{geometry}
\usepackage{setspace}
\usepackage[acronym]{glossaries}
\begin{document}

\author{Aleteia Araujo \and Breno Costa \and João Bachiega Jr. \and Leonardo R. Carvalho \and Rajkumar Buyya}
\textbf{Aleteia Araujo \{1,2\}, Breno Costa\{1\}, João Bachiega Jr.\{1\}, Leonardo R. Carvalho\{1\}, Rajkumar Buyya\{2\}}
\\\\
\{1\}Department of Computer Science\\University of Bras\'{i}lia (UnB)\\ Brasília - DF - Brazil \\ \textbf{email}: aleteia@unb.br and 
\{brenogscosta,joao.bachiega.jr,leouesb\}@gmail.com \\\\
\{2\} Cloud Computing and Distributed Systems (CLOUDS) Labs\\
School of Computing and Information Systems \\
The University of Melbourne, Australia  \\
 \textbf{email}:\{rbuyya@unimelb.edu.au\}
\\\\

\let\originalcleardoublepage\cleardoublepage
\renewcommand{\cleardoublepage}{\clearpage}
\tableofcontents
\let\cleardoublepage\originalcleardoublepage

\chapter{Observability in Fog Computing}

\textbf{Abstract} - 
Fog Computing provides computational resources close to the end user, supporting low-latency and high-bandwidth communications. It supports IoT applications, enabling real-time data processing, analytics, and decision-making at the edge of the network. However, the high distribution of its constituent nodes and resource-restricted devices interconnected by heterogeneous and unreliable networks makes it challenging to execute service maintenance and troubleshooting, increasing the time to restore the application after failures and not guaranteeing the service level agreements. In such a scenario, increasing the observability of Fog applications and services may speed up troubleshooting and increase their availability. An \hl{observability}
\index{observability}%
system is a data-intensive service, and Fog Computing could have its nodes and channels saturated with an additional load. In this work, we detail the three pillars of observability (metrics, log, and traces), discuss the challenges, and clarify the approaches for increasing the observability of services in Fog environments.
Furthermore, the system architecture that supports observability in Fog, related tools, and technologies are presented, providing a comprehensive discussion on this subject. An example of a solution shows how a real-world application can benefit from increased observability in this environment. Finally, there is a discussion about the future directions of Fog observability. 

\textbf{Keywords} - 
Observability,  Fog Computing, Edge Computing, Metrics, Logs, Traces
\\
\begin{table}[h]
    \centering
    \begin{tabular}{ll}
        \hline
        \textbf{Abbreviation} & \textbf{Meaning} \\
        \hline
        CNCF & Cloud Native Computing Foundation \\
        eBPF & enhanced Berkeley Packet Filter \\
        GCT & Garbage Collection Truck \\
        IoT & Internet of Things \\
        MEC & Mobile Edge Computing \\
        ODLC & Observability Data Life Cycle \\
        OTel & OpenTelemetry \\
        P2P & Peer-to-Peer \\
        SLA & Service Level Agreement \\
        SNMP & Simple Network Management Protocol \\
        SPOF & Single Point of Failure \\
        TSDB & Time Series Database \\
        WMI & Windows Management Instrumentation \\
        \hline
    \end{tabular}
    \label{tab:abbreviations}
\end{table}

\section{Introduction}
\hl{Fog Computing} 
\index{Fog Computing}%
is a model that seeks to bring processing and data storage closer to the location where they are required, usually at the network's edge. The primary features of Fog Computing include its distributed and decentralized nature, the emphasis on low latency and high bandwidth communication, and its capacity to support diverse devices and applications ~\cite{Bonomi2012}. These features make it particularly well-suited for delivering high Quality of Service (QoS) in applications that demand strict performance requirements. Additionally, Fog Computing can utilize network infrastructure and resources, such as routers, gateways, and edge devices, to create a more efficient and cost-effective computing environment~\cite{iorga2018fog}.

In the realm of \hl{IoT applications},
\index{IoT applications}%
Fog Computing plays a vital role in facilitating real-time data analytics, processing, and decision-making at the network edge. This proves to be essential for applications that demand low latency and high-bandwidth communication, such as industrial automation, autonomous vehicles, and smart cities~\cite{Naha2018}. Thus, Fog Computing enhances the intelligence and autonomy of IoT systems, alleviates network congestion and latency, and improves the overall reliability and performance of IoT applications~\cite{Naha2018}.

IoT applications often possess a more fragmented code base in comparison to traditional applications. This fragmentation arises because they typically involve numerous devices and systems that need to cooperate smoothly, necessitating various programming languages, frameworks, and protocols. Consequently, this scenario adds complexity to system maintenance and elevates the risk of failing to meet \hl{Service Level Agreements} 
\index{Service Level Agreements}%
(\hl{SLA})
\index{SLA}%
~\cite{vaquero2019research}. One potential solution to mitigate these challenges is to increase the observability of IoT applications, allowing timely decision-making and rapid troubleshooting of applications\cite{pallewatta2023placement}.

The concept of observability originates from control theory and has recently been utilized in the monitoring and debugging of distributed systems \cite{karumuri2021towards}. It denotes the ability to deduce the internal state of a system through its external outputs \cite{kalman1960general}. \hl{Monitoring}
\index{Monitoring}
involves collecting and evaluating performance metrics from applications to ensure that SLAs are met \cite{srirama2023decade}. Observability is seen as an extension of monitoring, employing data analytics techniques on collected monitoring data to accelerate the identification of the reasons behind system malfunctions \cite{marie2019demonstration}. In addition to metrics, observability incorporates application logs and service call traces, facilitating cross-examination of these data types to gain deeper insights into system behavior \cite{karumuri2021towards}. This approach offers better avenues for system intervention and minimizes the time needed to restore applications to a stable condition. Nevertheless, an observability platform must efficiently handle and process substantial data volumes promptly, while Fog Computing consists of resource-constrained nodes linked by potentially unreliable networks. Resource management in Fog Computing remains a developing field of research \cite{Mukherjee2024}, and this significant challenge needs further investigation \cite{usman2022survey}. 

We systematically reviewed the literature on Fog monitoring solutions in our previous work \cite{costa2022monitoring}. We have found that no available proposal could adequately manage the collection of metrics, logs, and traces simultaneously in a Fog Computing environment. This work clarifies the approaches used to increase observability in Fog environments, the system architecture that supports it, the related tools and technologies, and the open challenges of this field in this scenario. 

This work is organized as follows.  Section \ref{sec:background} describes the concepts that are the basis for the rest of the work. Section \ref{sec:challenges} presents the challenges and approaches related to observability in Fog Computing. The architecture of an observability system is described in Section \ref{sec:architecture} . Section \ref{sec:tools} discusses the tools and technologies that support an observability system in Fog environments. Section \ref{sec:solution} shows an example of an observability solution.  The future directions are discussed in Section \ref{sec:future}. Finally, Section \ref{sec:summary} concludes this chapter.

\section{Background}
\label{sec:background}

This section summarizes the relevant characteristics and architecture of Fog Computing. It also details observability in the context of distributed applications and presents the life cycle of observability data on Fog environments. 

\subsection{Fog Computing} 
\label{Subsec:Fog}

The concept of Fog computing was presented in 2012 by Cisco \cite{Bonomi2012} to address the challenges of IoT applications running in Cloud Computing environments. Yi et al. \cite{Yi2015} consider Fog Computing to be made up of a large number of decentralized and heterogeneous devices, where they communicate and cooperate among themselves and with the network to perform data processing and storage without third-party intervention. Fog Computing is a distributed computing paradigm that acts as an intermediate layer between Cloud datacenters and \hl{IoT devices},
\index{IoT devices}%
overcoming the high latency characteristic of the Cloud by using idle resources of various devices near end users \cite{Mahmud16,Naha2018}.

The six characteristics of Fog Computing, defined by NIST \cite{iorga2018fog}, are crucial for enabling a more efficient and scalable Fog Computing environment. These characteristics collectively enhance the capability of Fog Computing to process and analyze data closer to the source, reducing latency and improving the speed of decision-making processes. These six essential characteristics of a Fog Computing environment are defined as follows:
\begin{itemize}
	\item \hl{Low latency}:
 \index{Low latency}%
  by knowing the logical location of IoT devices and being in their vicinity;
	\item Geographical distribution: allows that a service or application be widely distributed physically;
	\item \hl{Heterogeneity}:
 \index{Heterogeneity}%
  devices with different computational capacities and with heterogeneous software stacks cooperate using multiple types of networks;
	\item Interoperability: a service can span by the infrastructure of different providers;
	\item Real-time interactions: applications of Fog Computing involve real-time interactions rather than batch processing;
	\item Scalability: must provide adaptability and scalability, use resource pooling, and survive data-load changes and network condition variations.
\end{itemize}

Fog Computing is especially important for applications that require real-time responses, such as autonomous vehicles, smart cities, and industrial IoT. 
Additionally, the support for mobility and geographical distribution ensures that data can be processed and acted upon regardless of location. At the same time, heterogeneity and interoperability allow seamless integration across various devices and systems. These attributes make Fog Computing a versatile and powerful complement to traditional cloud computing, driving innovation in multiple industries.

There are other similar paradigms, such as \hl{Edge Computing}~\cite{Dastjerdi2016},
\index{Edge Computing}%
 Mobile Edge Computing (MEC) \cite{Dolui2017}, and Mist Computing \cite{ranaweera2021survey}, which are frequently confused with Fog.  In this chapter, we consider Fog Computing as a broader and more complete concept that can be regarded as an umbrella that encompasses all other similar paradigms \cite{chiang2017clarifying}. In a broad sense, this means that the solutions proposed for one of these similar paradigms may be able to run in the others after some level of adaptation.

\begin{figure}[h]
	\centering
\includegraphics[width=0.65\textwidth]{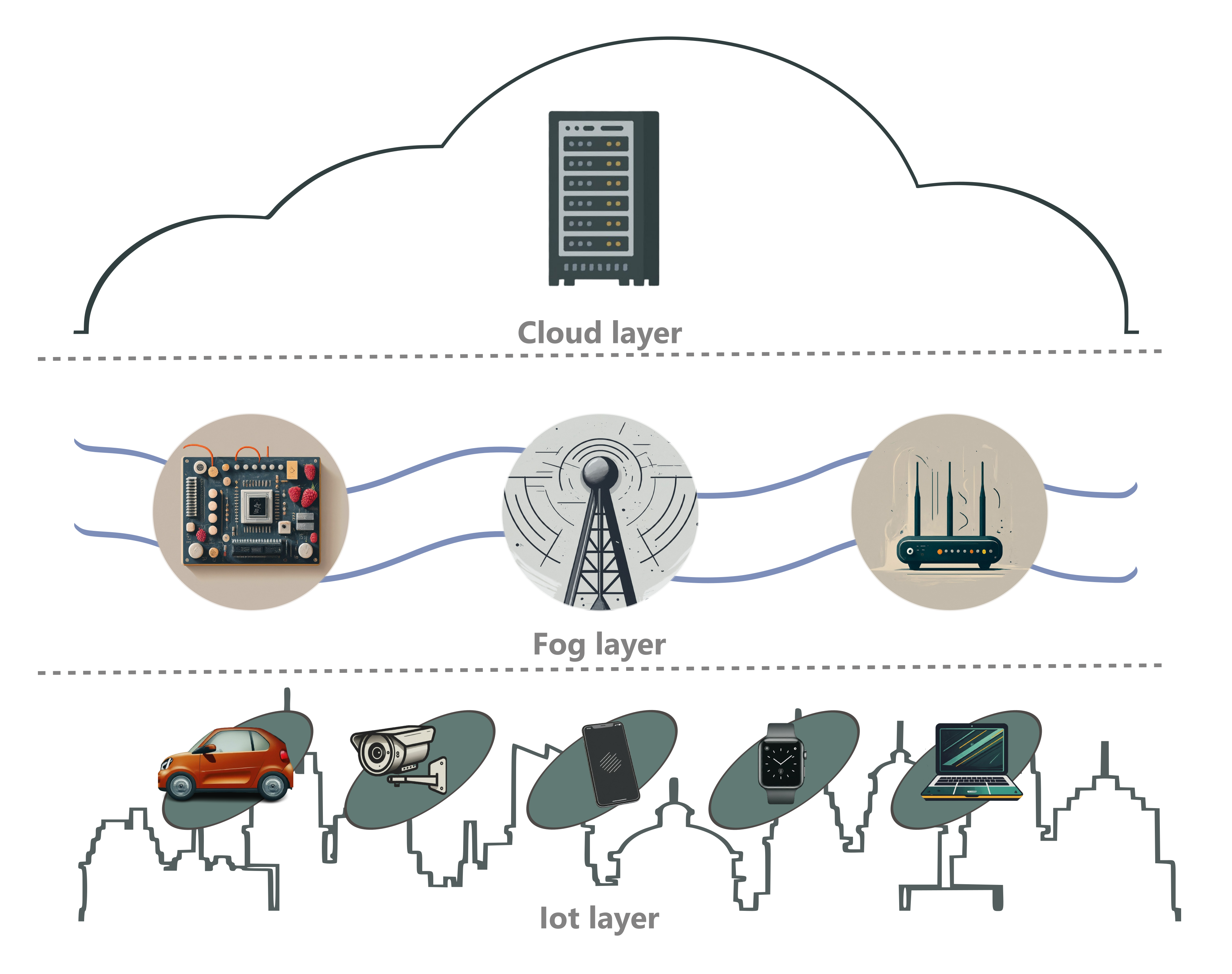}
	\caption{Overview of the Fog Computing architecture.}
	\label{fig:fogoverview}
\end{figure}

The architecture most commonly used to represent a Fog Computing environment comprises three layers: IoT layer, Fog Layer, and Cloud Layer, as presented in Fig. \ref{fig:fogoverview}.
The \hl{IoT layer}
\index{IoT Layer}%
is made up of IoT devices connected at the edge of the network, through which end-users can request services to be processed in the above layers. For example, users can request a list of restaurants whose noise level has been below 70 dB in the last 15 minutes. 

\begin{figure}[ht]
	\centering
\includegraphics[width=0.75\textwidth]{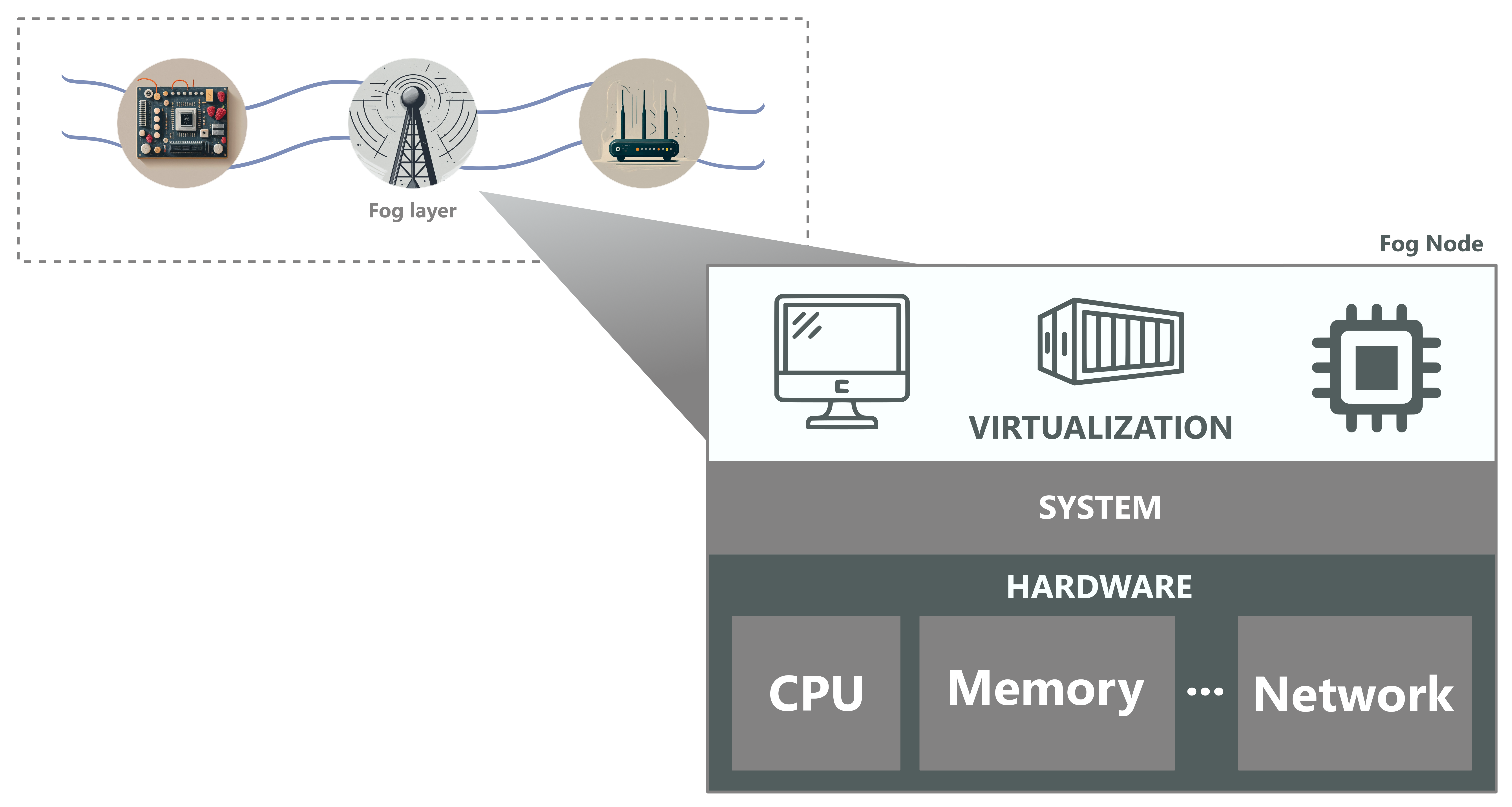}
	\caption{Resource abstraction on a Fog node.}
	\label{fig:node}
\end{figure}

The \hl{Fog Layer}
\index{Fog Layer}%
is located between the IoT and Cloud Layers. Provides shared resources that IoT applications can use as needed, such as processing and data storage resources, before data are transferred to the Cloud \cite{AlDoghman2016}. 
This layer consists of nodes, commonly called \textit{Fog nodes}. A Fog node is any device in a Fog Computing environment that can share spare resources with IoT devices and other Fog nodes. Many devices can be a Fog node, including smartphones, routers, notebooks, and specialized servers. The \hl{Fog node}
\index{Fog node}%
comprises physical resources (e.g., CPU, memory, network interface, among others) and system resources necessary for the abstraction of hardware and the execution of applications. A common way to overcome hardware and system heterogeneity is using a virtualized environment such as containers, as shown in Figure \ref{fig:node}. A comprehensive analysis of the computational perspective of a Fog node can be found in \cite{bachiegafognode}.
Finally, the Cloud Layer comprises Cloud providers' services, with more robust computational resources to deliver high-order processing and long-term storage.
The existence of the Cloud is fundamental in a Fog environment \cite{AlDoghman2016}, because Fog Computing complements Cloud Computing, but does not replace it.

Thus, a Fog Computing environment that supports IoT applications is characterized by having a more distributed organization, heterogeneity of physical devices and networks, and uncertainty of connectivity caused by device mobility, network instabilities, and battery exhaustion \cite{iorga2018fog}. This scenario differs from a Cloud Computing environment, supported by homogeneous resource-rich servers, continuous power supply, and stable redundant network connections.

\subsection{Observability}
\label{subsec:Observability}

\hl{Observability}
\index{Observability}%
is a characteristic of systems that provide information about their internal states using external output\cite{kalman1960general}. The higher the observability level, the easier it will be to understand the current and past behaviors of the system. From the time it is available, this knowledge allows for proper actuation of the system when needed. 

In the context of Fog Computing, observability is crucial for maintaining desired QoS levels and meeting SLA obligations. By providing real-time visibility into the performance of Fog services, observability enables operators to proactively manage QoS and take corrective action to prevent SLA violations \cite{brogi2017qos,haghi2020quality,hussain2014maintaining}. Observability tools continuously track key QoS metrics such as:
\begin{itemize}
    \item Bandwidth: The capacity of a network connection to transmit data within a given time frame. Adequate bandwidth ensures smooth data flow and prevents bottlenecks.
    \item Latency: The delay in data transmission, a critical factor in Fog applications that require real-time responsiveness. 
    \item Throughput: The rate of data delivery, reflecting the system's capacity to process workloads efficiently. 
    \item Uptime: The availability of Fog services, indicating the service's reliability and adherence to SLA uptime guarantees.
\end{itemize}
 By monitoring these metrics, observability tools can trigger alerts when QoS deteriorates, allowing operators to address performance issues before they impact end-users or lead to SLA breaches. 

Observability can be instrumented on a service or an application. There are three instrumentation domains of observability: \hl{metrics}, \hl{logs}, and \hl{traces}~\cite{karumuri2021towards}, 
\index{metrics}%
\index{logs}%
\index{traces}%
also known as the \hl{three pillars of observability}~\cite{sridharan2018distributed}.
\index{three pillars of observability}%
Each \hl{instrumentation domain}
\index{instrumentation domain}%
contributes to the observability of a system in different ways. Monitoring is traditionally related to the management of performance metrics, i.e., \% of CPU usage, \% of memory available, etc. Observability can be seen as a superset of monitoring in that it considers the collection, storage, and transmission of metrics and the management of other data types. Therefore, evolving from a traditional monitoring system to an observability system means not only an additional volume of data to be managed. It is necessary to deal with heterogeneity in data types, which will require different storage technologies and query requirements \cite{karumuri2021towards}, as well as other opportunities to optimize the use of computational resources. 

Dealing with different instrumentation domains will provide additional information and complementary views about the internal functioning of distributed systems. In this scenario, the strategies proposed and validated in the literature for traditional monitoring remain valid and can be fully incorporated into the context of observability. In addition, algorithms and approaches that deal specifically with other instrumentation domains besides metrics, such as logs and traces, are necessary. Whenever possible, another good alternative is to generalize the common aspects of all instrumentation domains and deal with them uniformly.


Metrics, logs, and traces are the most critical instrumentation domains, according to the monitoring and observability literature \cite{karumuri2021cloud,marie2021observability}. They carry different types of information, and each of them independently contributes to increasing the observability of a system, allowing complementary actuation.

\textbf{Metrics} are more related to the performance of a system. They are numerical values collected at a point in time, and their collection can be characterized as a time series. In the scenario of an IoT application, the following IoT device metrics can be available: percentage of CPU usage,  speed (if a vehicle), throughput of the 5G network in Mbps, amount of data transmitted in MB, etc. Different metrics may lead to various actuations. A network throughput that has dropped below a given threshold may signal that data transfers should be postponed to another moment when the throughput is higher.
As it concerns the system's external interface, metrics can be collected by a probe or agent outside the system, deployed on the same device, or accessed by the network. Data volume is generally low and steady and is directly proportional to the number of metrics collected multiplied by the collection frequency.  Metrics are commonly used in two ways: i. to generate alerts, such as when CPU usage reaches 90\%; and ii. to analyze and dashboard, such as showing in a graph the city areas where 5G network throughput is on average bellow of 50Mbps in the last month.

\textbf{Logs} are unstructured or semi-structured text files reporting relevant events and contextual information, whose instrumentation is usually made at development time, e.g., the developer decided to write successful transactions and error details in the log. Using the Smart City scenario, application logs can have information related to the QoS of the network connection, the geographic coordinates of the device, etc. Analyzing those data over time makes it possible to determine locations where network throughput is low at many moments of the day. This insight may lead to an investigation of the coverage of the 5G network that can be sent to the network carrier.
Metrics deliver objective information about a system's external interface, e.g., data upload throughput. They enable fast decision-making in response to measurements outside a specified regular range. The frequency of metrics collection runs in tandem with the required speed of decision-making, creating a trade-off between information timeliness and the resource consumption of running data processing and transmission tasks. 
Logs usually provide internal information about failure events, such as specific error messages, exception handling messages, and runtime errors. They can help speed up root cause analysis, help maintenance staff improve error treatment, and return the system to a healthy state. 
However, the relevant information in the logs must be extracted to separate the wheat from the chaff. Usually, a practical log analysis may need to assess logs written a few hours or even days before the abnormal event is examined. 

\textbf{Traces} are records of service calls made by the system. They allow observations of the service call sequence and the time spent on each call from the beginning to the end of a request. Trace analysis can show which service calls are taking longer in the response time composition of an application. They can also show requests that do not finish properly. 
In the former case, the action could be code optimization delivered as a new application version. In the latter case, better error management can make the application more resilient to incomplete request processing. In the Smart City scenario, the application performance information (data upload throughput) can be aggregated by suburb of the City. 
The traces supply details about the internal flow of information, including the sequence and delay of each service call needed to process a request. These data can be visualized as a graph and a critical path can be generated from it, allowing for scrutiny of the dependency among the components of a system \cite{sigelman2010dapper}. The volume of data by period depends on the number of requests and can be sporadic, reaching its peak while the system is experiencing peak demand.

Connecting the three domains by the time each information is generated is feasible. 
When it is viable to relate two or three of them in the same analysis, more opportunities for actuation arise. Log data, when collated with metrics collected during the same period, provide a more comprehensive inspection of runtime issues by simultaneously compiling external and internal views of the system. A trace can be seen as a breakdown of a response time metric, allowing identification of the components where an improvement in processing or communication delay can result in a shorter final response time. 

Having more instrumentation domains available means a higher {observability level}.
\index{observability level}%
In addition to the independent value of each domain, there is an additional value in cross-analysis between domains due to their \hl{synergistic interactions}~\cite{mcgrane2011method},
\index{synergistic interactions}%
i.e. when two or more factors act as causes of a particular outcome. This effect is popularly known as ``the whole is more than the sum of its parts''. 
We defined an observability index in a previous study \cite{costa_achieving_2024}. However, instead of having a formula where the observability level of a system is determined only by the number of instrumentation domains whose data are available for analysis, as in Equation \ref{equ:1}:

\begin{equation}
Observability = Metrics + Logs + Traces
\label{equ:1}
\end{equation}where 
\begin{equation}
  Metrics, Logs, Traces=\begin{cases}
    1, & \text{if the instrumentation domain data are available for analysis}.\\
    0, & \text{otherwise}.
  \end{cases}
\label{equ:2}
\end{equation}
We added the synergistic interactions between them as well, as in Equation \ref{equ:3}:
\begin{equation}
Observability = Metrics + Logs + Traces +  (Metrics X Logs X Traces)
\label{equ:3}
\end{equation}where X is an operator that filters the data for each different pair of the instrumentation domains (ID) available and returns the subset of each ID that matches a specific period. The expression (IDiXIDjXIDk) is the same as (IDixIDj)+(IDixIDk)+(IDjXIDk). And (IDiXIDj) may assume one of the following values:
\begin{equation}
    IDiXIDj=\begin{cases}
    1, & \text{if both Instrumentation Domains are equal to 1 (Equation \ref{equ:2}), AND}\\
       & \text{the applied time filter returns a nonempty set for both}.\\
    0, & \text{otherwise}.
  \end{cases}
\label{equ:4}
\end{equation}
This definition shows that to increase the observability of a system, it is important not only to collect the information from the instrumentation domains and to analyze each data set isolatedly. It is also relevant to be prepared to learn from their interactions to generate more value from the same available data sets.

\subsection{Fog Observability Data Life Cycle}
\label{subsec:cycle}
\begin{figure*}[ht]
	\centering
	\includegraphics[width=\textwidth]{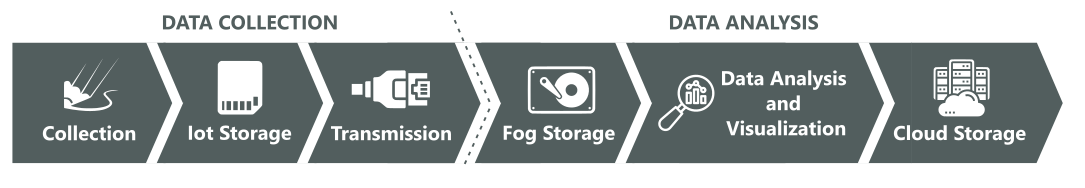}
	\caption{Fog observability's data life cycle.}
	\label{Fig:LifeCycle}
\end{figure*}

\index{Observability Data Life Cycle}%
To obtain valuable information from each instrumentation domain and to increase the observability of an application running in a Fog environment, it is necessary to be aware of the following six-step \hl{Observability Data Life Cycle} (ODLC)\cite{costa_achieving_2024}, depicted in Fig. \ref{Fig:LifeCycle}:  1. Collection; 2. IoT storage; 3. Data are transmitted to Fog; 4. Fog storage; 5. Data analysis and visualization; 6.Cloud storage and analysis. The first three steps comprise the Data Collection phase of the life cycle. The last three steps form the Data Analysis phase.

\textbf{1. Collection}

The data are collected in the initial step of the Fog observability data life cycle. This can happen in many ways depending on the instrumentation domain in place. Metrics can be acquired from the operating system by utilizing system calls that report the amount of available resources (e.g., CPU, memory, disk storage). 
Platforms such as container management systems usually also report resource usage metrics. Finally, the application can perform specific processing to report, for instance, the average response time or the number of requests delivered in a period.
\hl{Logs}
\index{Logs}%
are written according to the specific event flow instrumented to be recorded in text. For instance, successful events, such as HTTP requests answered with code 200, and failures, like function call stack, in the case of runtime exceptions caught in the code base.
When previously instrumented, specific API calls can create traces that record the sequence and delay of each service call. 

Computational resources such as CPU cycles and memory space must be used to collect observability data. The higher the number of metrics collected, rows written in a log file, and traces recorded, the higher the observability overhead added to the environment. A Fog environment is made up of resource-restricted devices. Therefore, the overhead caused by the observability of data collectors should be low.
 
\textbf{2. \hl{IoT Storage}}
\index{IoT Storage}%
In this step, the observability data were collected and now are in device storage awaiting transmission or removal. As these data are usually immutable and append-heavy \cite{karumuri2021cloud}, the trend is that their volume becomes more extensive over time. 
A data removal policy should be in place to avoid running out of storage resources. For example, every time data are transmitted to the Fog, it should be deleted to free up space for more data to be stored. This removal can be only partial if a minimum data window is necessary for some local processing. 
The period in which a device can handle the observability data stored will depend on several factors, such as the data footprint by period, the frequency of generation, and the available storage space reserved for the system. 

Although metrics can be stable with respect to data volume, logs and traces have more significant variability \cite{karumuri2021towards}. This characteristic makes it difficult to monitor and actuate when running out of space, which is risky. 
In addition to local data deletion, alternative actuation can be in place, such as signaling to the observability server that there is a risk of data loss and verifying if the dynamic circumstances allow those data to be immediately transmitted and erased. 

\textbf{3. Transmission}

Observability may allow for proper and timely decision-making. Although it is possible to make some minor decisions locally by a single device, critical decisions are expected to be made by a process that can assess a higher volume of data coming from different system subcomponents, granting a more comprehensive view of the system. 
Therefore, data collected from the IoT layer should be transmitted to the Fog Layer, where a resource-richer node will store them, allowing for a more comprehensive data analysis. 

Because \hl{network uncertainty}
\index{network uncertainty}%
is a characteristic of Fog Computing, a suitable \hl{Fog observability}
\index{Fog observability}%
solution must deal with temporary network outages. Ideally, it should adaptively modulate the observability data flow, considering the amount of data to transmit, their type (metrics, logs, traces), and the available bandwidth.
The network connections used by the application to receive and respond to user requests may be the same as those used by the observability data flow \cite{dastjerdi2016fog}. Thus, transmitting large volumes of observability data can strain the network, particularly in Fog environments with limited bandwidth. This can increase latency for both observability data and application traffic, negatively impacting QoS and potentially leading to SLA violations.
An adaptive process may be in place to define the amount of data that can be transferred from the devices, selecting which instrumentation domains will be included in each transmission and the period to which the collected data will refer.

\textbf{4. \hl{Fog Storage}}
\index{Fog Storage}%
Fog nodes are expected to be resource-richer compared to IoT devices \cite{bachiegafognode}. Due to this, it is on the Fog Layer where observability data from several IoT devices are stored for rapid actuation and decision-making. In this step, the received data can be preprocessed to get contextual information, e.g., the addition of device ID. However, Fog nodes are not as rich in resources as Cloud nodes \cite{bachiegafognode}. Therefore, the volume of data should be limited to a volume that these devices can handle.

Metrics can be seen as time series, and a \hl{time series database} (\hl{TSDB})
\index{time series database}%
\index{TSDB}%
should be used to store them optimally. However, logs and traces are structured differently and will benefit from other storage solutions. Karumuri et al. \cite{karumuri2021towards} analyzed observability data in a Cloud environment and considered that logs and traces would benefit from inverted index-based storage due to the type of queries usually made to retrieve meaningful information from them.
Therefore, an observability data ingestion service in Fog should consider the data requirements of each instrumentation domain (see Table \ref{tab:domains}), while allowing cross-analysis to be performed.


\begin{table}[!t]
\caption{The three domains of observability differ in their data characteristics \cite{karumuri2021towards}.}
\label{tab:domains}
\centering
\begin{tabular}{@{}llll@{}}
\toprule
\textbf{Domain} & \textbf{Type}                 & \textbf{Query}             & \textbf{Storage}     \\ \midrule
Metric          & Numeric                       & Aggregations               & Time Series Database \\
Log             & Semi/not structured strings       & Approximate string search & Inverted Index       \\
Trace & DAGs of the duration of execution & Disassociated graph search & Inverted Index  \\\bottomrule
\end{tabular}
\end{table}

\textbf{5. Data Analysis and Visualization}

Once the observability data are available on the Fog, it is possible to query them and make decisions and actions accordingly. In this step, the value of Fog Computing can be delivered. Located on the edge of the network, a Fog service can provide faster responses to IoT devices and end users. Additionally, being located between the IoT and the Cloud layers prevents congestion on the Cloud network.

Observability data tend to give more relevant answers when they are queried as soon as they arrive, which means that most queries and analyzes use more recent data (less than 24 hours) \cite{karumuri2021cloud}. Thus, it is essential to guarantee fast access to this time window. In addition, to save resources to continue receiving IoT data, it is crucial to provide automated mechanisms to send the data out of this range to long-term storage in the Cloud. 

\textbf{6. \hl{Cloud Storage}}
\index{Cloud Storage}%
The Cloud is the appropriate environment to store large data volumes and run heavy data processing models, such as historical analysis of observability data \cite{hashem2015rise}. 
Data can be automatically exported from the Fog storage system when they stay outside the pre-configured time window, e.g., a week after their collection time. These data might be moved to the Cloud, maintaining a predetermined volume of data on the Fog storage and helping to guarantee a low query response time.

As seen previously, the life cycle steps presented are the same for each instrumentation domain. However, they have different characteristics (see Table \ref{tab:domains}) and contribute differently to the system's observability, depending on the circumstances. For instance, in cases where collected metric values show suitable device performance but there are identified runtime issues, one can temporarily collect only logs and traces, aiming to speed up the discovery of the root cause of the problem while lowering system overhead related to observability. A resource-restricted environment like Fog might benefit from such adaptive behavior.

After presenting how Fog Computing and Observability are defined in the literature and the steps necessary to make observability data flow from the IoT Layer to the Cloud Layer, the following section describes the approaches to implementing observability systems in distributed environments and the challenges that should be overcome.

\section{Challenges and Approaches in Fog Observability}
\label{sec:challenges}
The \hl{observability of distributed systems}
\index{observability of distributed systems}%
has become essential to maintain stability, performance, and security. It enables organizations to identify and resolve issues proactively, optimize resource utilization, and ensure a smooth user experience.  In the context of Fog Computing, these capabilities are particularly critical for managing QoS and ensuring SLA compliance. However, some challenges should be considered when designing and implementing a Fog Observability system. This section will present some of these challenges and discuss the approaches observability systems use in this environment.

\subsection{Challenges}

Defining open scientific challenges in observability on Fog Computing is crucial for advancing the field and addressing the complexities inherent to distributed computing environments. Observability involves monitoring and gaining insight into systems' performance, reliability, and security, which is incredibly challenging in fog environments due to their decentralized nature and the diversity of devices involved. By identifying and understanding these challenges, researchers and practitioners can develop more effective tools and methodologies to monitor and manage fog networks, ensuring optimal performance and resilience.

Furthermore, addressing these challenges is essential for improving the reliability of critical applications, such as those in healthcare, autonomous systems, and smart infrastructure, where the ability to observe and respond to system behavior in real-time can have significant implications. Ultimately, a focused exploration of these open challenges will create more robust, scalable, and secure \hl{Fog Computing}
\index{Fog Computing}%
solutions. 
The main open challenges are: 
\begin{itemize}
    \item Heterogeneity of devices, a Fog characteristic; 
    \item Diverse and separate data sources \cite{kosinska2023toward};
    \item  Resource consumption of  observability system's components and ODLC \cite{levin2020viperprobe}; and 
    \item Security \cite{petrakis2018internet}.
\end{itemize}

\subsubsection{\hl{Heterogeneity} of devices} - 
\index{Heterogeneity}%
The variety of capabilities, network protocols, applications, and vendors among IoT devices and sensors, coupled with the unique characteristics of each IoT environment, presents significant challenges in the deployment of observability systems \cite{fizza2021qoe}.  The diversity in the hardware of IoT devices may result in software stacks that are not universally compatible across all devices \cite{babu2021fog}. And an usual observability approach considers the use of software agents deployed in the devices. This can lead to a significant development challenge in managing multiple versions of the same software, with the added complication that these versions may not provide identical functionality due to the variations in operating systems and in other platforms that support them. Using containers as an execution environment can reduce the risk of incompatibility between the IoT component of an observability system and the software stack of the hosting device. Eventually, there must be compatibility between the device and the container platform. 

\subsubsection{Diverse and separate data sources} -
Increasing the observability of services and applications means collecting metrics, logs, and traces and correlating them comprehensively, as we can see in Equation \ref{equ:3}. It involves collecting data that may be difficult to consolidate and relate because they originate from various sources, have distinct types, and address different concerns. In addition, resource restriction of IoT devices, added to the connection uncertainty and characteristics of Fog Computing, increases the risk of observability data loss. Even when having a stable data connection, resource exhaustion (in devices, in the network, or in the server) can initiate actuations that reduce the collection of observability data to a fraction of what is generated. For example, a strategy to minimize resource consumption in distributed tracing collection is to sample the data, collecting only one trace out of 10,000 generated traces \cite{gatev2021introducing}.
These factors challenge the data model and its analysis, which requires the construction of a comprehensive view of these diverse and fragmented perspectives \cite{kosinska2023toward}. 

\subsubsection{Resource consumption of  \hl{observability system}'s components and ODLC} -
\index{observability system}%
Another challenge of Fog observability systems is appropriately managing a large data volume in a dynamic and potentially restricted environment \cite{karumuri2021towards}. Fog infrastructure, network channels, and applications are instrumented to generate large volumes of heterogeneous time series data that must be transmitted, indexed, stored, and queried in near-real time, and this will ingest an additional load on an already saturated system\cite{karumuri2021towards}. Inefficient data mobility may overload a Fog node, affecting service level agreements \cite{jonathan2017nebula}. Pallewatta et al. \cite{pallewatta2023placement} suggest incorporating data mining and artificial intelligence models with a microservice placement strategy to make performance-aware decisions and address anomalies and failures in the system. Petrakis et al. \cite{petrakis2018internet} proposed a framework that optimally handles sensor data with minimal bandwidth usage. Their approach includes data filtering to reduce communication overhead, local data caching for on-site data processing, deferred data uploading to deal with limited or no-bandwidth situations, and, ultimately, synchronization with Cloud-based data. Another approach to prevent the high volume of observability data from negatively impacting the domain application in Fog \cite{Dastjerdi2016} is to provide self-adaptable behavior to the observability system. Depending on the overall load on the system, be it on the end-user device or the Fog node, certain decisions can be made to reduce the impact of the observability data flow. This adaptive behavior can be applied to the frequency of data collection, the volume and type (instrumentation domain) of data to be transmitted, and the communication model. The aim is to limit or delay specific actions when the system load is high and try to resume them as soon as they return to a normal state \cite{costa2022monitoring}.  In addition, a most seminal challenge in terms of setup is bootstrapping \cite{fersi2021fog}. Bootstrapping is related to the installation of the observability agent in the IoT device for the first time and the amount of time and battery that this process consumes.

\subsubsection{Security} -
\hl{Security}
\index{Security}%
is an open challenge in collecting, storing, and transmitting observability data in the Fog environment. Fog nodes can be distributed widely by a large environment, making it challenging to enforce physical security policies as it is usual in on-premise environments \cite{babu2021fog}. In addition, the resource restriction of many IoT devices and sensors hardly limits the security options that require high computational processing, such as cryptography. However, observability data can contain critical information about the status and localization of devices, such as vehicle speed and geographic coordinates, that can cause a privacy breach if exposed. In addition, client software can often be easily compromised \cite{petrakis2018internet}. Due to these privacy and security risks and the proximity of IoT devices to end users, Fog nodes must initially implement access control, encryption, contextual integrity, and isolation mechanisms to protect sensitive data before they leave the node. 

\subsection{Approaches}

The evolution of observability approaches has become increasingly crucial in Fog Computing. Traditional monitoring tools, which focus primarily on tracking predefined metrics or logs, are no longer sufficient to capture the intricate behaviors of modern systems. As a result, observability has evolved to encompass a more holistic view, integrating metrics, logs, and traces to provide comprehensive insight into system performance and health. In this way, each method has unique features, benefits, and drawbacks \cite{kufel2016tools}. This section analyzes these approaches, shedding light on their strengths and limitations.

A popular approach is \hl{agent-based observability},
\index{agent-based observability}%
which involves installing software on each monitored system to collect detailed data. These agents are domain-specific and can capture various metrics, such as CPU utilization, disk utilization, and application activity. This approach provides deep monitoring and enables rapid issue detection. However, relying on agents introduces scalability challenges, as deploying and managing agents in large environments can be complex. In addition, agents can be resource intensive, potentially affecting the performance of monitored systems.

In contrast, \hl{agentless observability} 
\index{agentless observability}%
uses the system's built-in monitoring technologies and protocols, such as Windows Management Instrumentation (WMI) and Simple Network Management Protocol (SNMP). Eliminates the need for additional software and offers simplified deployment and management. This approach is particularly well-suited for monitoring system availability and basic performance metrics. However, agentless observability often lacks the granularity and depth of data that agent-based approaches offer. Organizations looking to combine the strengths of both approaches may opt for a hybrid approach. This method allows flexibility in choosing between agent-based and agentless methods based on the specific requirements of the monitored systems. Critical systems can employ agents for granular data collection, while standard systems can be monitored using agentless methods to minimize overhead.

With the rise of Cloud environments and complex distributed systems, data streaming approaches have gained prominence \cite{kufel2016tools}. This approach involves integrating data forwarders into the application code, allowing performance metrics to be transmitted as streams to a central observability server. By capturing application-level data, operations teams can gain real-time insight into application behavior, user experience, and performance bottlenecks. However, implementing data forwarders requires code modifications and can introduce overhead.

Choosing the most appropriate \hl{observability approach}
\index{observability approach}%
depends on factors such as the organization’s requirements, resource constraints, and the type and scale of the monitored systems. Organizations should carefully consider the pros and cons of each approach to make an informed decision that aligns with their business objectives \cite{kufel2016tools}.

\section{System Architecture}
\label{sec:architecture}

An observability system provides the components and functions necessary to monitor the environment (sensors, devices, operating systems, networks, and applications), collect instrumentation data (metrics, logs, and traces) and store them to allow automated actuation and troubleshooting and maintenance of services and applications. The system can show an updated status of the monitored assets and make it easier to investigate and correct failures and other issues that usually arise in a dynamic and complex environment. Due to some Fog characteristics, such as resource restriction, device and node mobility, and unreliable communication channels (described in Section \ref{Subsec:Fog}), the collection, storage, and analysis of a large volume of data in Fog Computing remains an open challenge. This section presents the software architecture of an observability system in a Fog Computing environment.

\subsection{Architecture}
\label{subsec:architecture}

An \hl{observability system}
\index{observability system}%
comprises the same structural functions as a traditional monitoring system: i. observation of monitored resources and services; ii. data processing and analysis; and iii. data exposition \cite{abderrahim2017holistic,brandon2018fmone}. Observation means the acquisition of updated statuses of resource usage (e.g., CPU load and network latency) or service performance (e.g., execution logs and call traces). Processing and analysis are related to the necessary adjustments and transformations required on the data, such as data filtering and aggregation, creating and managing events, and notifications derived from pre-configured rules and thresholds. Exposition is related to where the generated data are stored (e.g., in a local database, JSON files) and how it can be accessed by operational and maintenance staff (e.g., visualization through dashboards and other functionalities consuming the data directly in the storage).

The monitored subject must provide access to its metric values, logs, and traces. This is usually implemented using an operating system or service calls made by a local agent that collects the data in a predefined recurrent period and makes them available for transmission. After transmission, observability data can be processed by a specialized component and transformed in some way (context addition, data aggregation or summarization, etc.) according to the service requirements before they are stored. After storing the data, it is possible to verify whether a predefined threshold was reached or if an anomaly was detected. If so, proper actions can be taken, such as creating an event or notifying maintenance personnel. Finally, the observability server is responsible for storing collected data from all distributed subjects and making them available for queries and visualization. 

A generic architecture to deliver the functionalities described in the last paragraphs is composed of three logically independent components, namely: 
\begin{itemize}
    \item an \hl{Agent} component that runs in every device that will be monitored; 
    \item a \hl{Transformer} component that can aggregate and transform the data before sending it to the server, and;
    \item a \hl{Server} component, responsible for providing a comprehensive view of observability data to the users and other systems.
\end{itemize}
\index{Agent}%
\index{Transformer}%
\index{Server}%
The Agent component can be implemented as software that collects observability data from heterogeneous devices such as Small Board Computers (e.g., Raspberry Pi), Smartphones, edge devices, and Cloud servers. Metrics related to device resource consumption, such as the percentage of CPU used, the percentage of storage available, logs of a container environment, or the status of a running service, can be collected. The data is stored in the device and can trigger local events based on configured thresholds. The Agent component sends data to a Transformer or a Server component that subscribed previously and deleted the oldest data collected whenever it needs room to store more recent data. The frequency of collection, communication mode (e.g., pull, push), and other component subscriptions are parameters that the Server component can dynamically modify.

The Transformer component receives observability data from several Agents. Its primary role is to absorb the data processing burden from devices near it (regarding the number of hops or latency). The data can be aggregated by period, location, application, or transformed in some way before being sent to the Server. Although Agents usually run on resource-restricted and heterogeneous devices, the Transformer component needs to run on a device with enough resources to process and store more data. Its presence in the communication flow is not mandatory. If the observability data collected from a specific device does not require intermediary processing or storage, it can be delivered directly to the Server.

The Server component receives all observability data from the monitored devices and services. Provides an interface that allows data visualization for operations and maintenance staff, and shares its database with other modules. The data can be used for historical analysis, forecasting, root-cause analysis, etc. The Server is aware of the status of all components of the observability system. It can modify their execution parameters remotely, limiting or stopping observability data collection for some time. This is useful when a device is experiencing a lack of resources, f.i. when a device is running out of disk space. 

\begin{figure}
    \centering
    \includegraphics[width=0.75\linewidth]{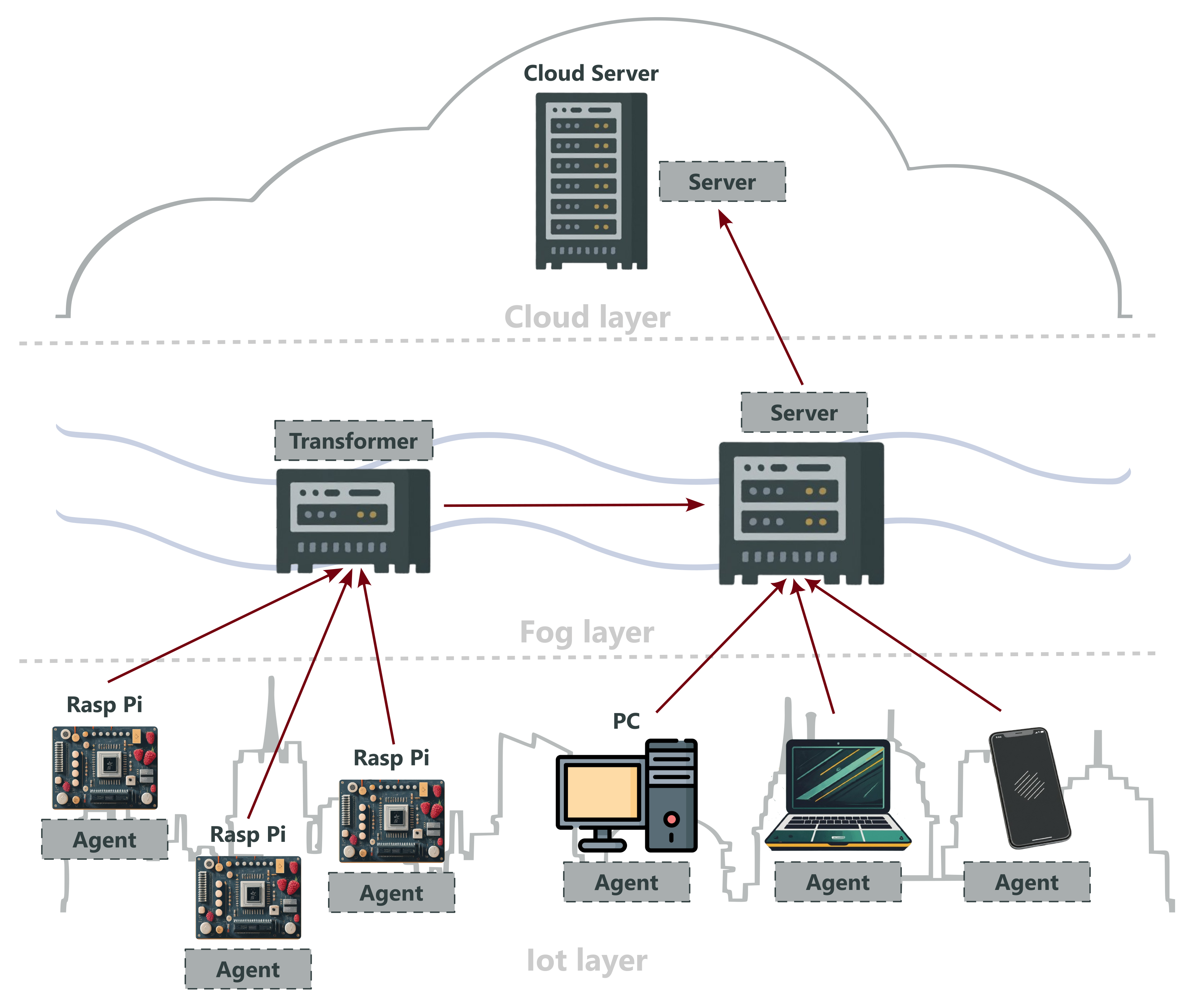}
    \caption{Components of a Fog Observability System.}
    \label{fig:architecture}
\end{figure}

Although described as three independent components, they can also be located in the same execution environment and implemented as different functions of the same service. Figure \ref{fig:architecture} shows how these components can be distributed considering a three-layer Fog architecture. On the left side of the Figure is a configuration where the Transformer component was necessary to aggregate and summarize observability data before sending them to the Server.

\subsection{Topology}
\begin{figure*}[ht]
	\centering
	\includegraphics[width=\textwidth]{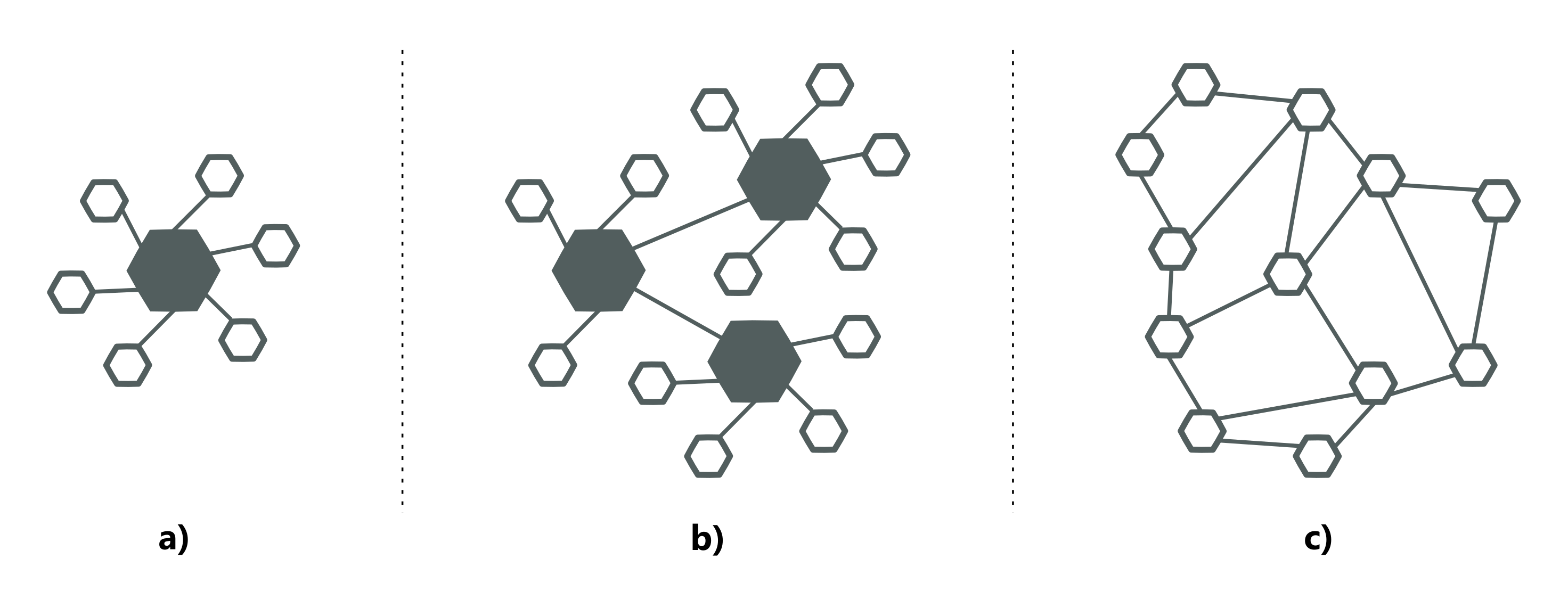}
	\caption{Topologies of observability systems: a) Centralized; b) Decentralized; c) Distributed.}
	\label{fig:topology}
\end{figure*}

The topology describes how the observability system is structured regarding the distribution of its components and data flow. Masip et al. \cite{masip2020collaborative} described three control topologies, as shown in Figure \ref{fig:topology}, that can be used in Fog environments and were adapted to observability systems' topologies: 
\begin{itemize}
    \item \textbf{\hl{Centralized}} -  
    \index{Centralized}%
    with only one observability server receiving data from several monitored sources (devices, services, and applications); 
    \item \textbf{\hl{Decentralized}} - 
    \index{Decentralized}%
    where a set of nodes have a local observability server, and these servers act somehow on the data (filtering, storing, etc.) and collaborate among them to compose a complete view of monitored sources; 
    \item \textbf{\hl{Distributed}} - 
    \index{Distributed}%
    where an observability component is localized on each Fog node, and all components interact to share their view of monitored resources and maintain the view of the whole environment updated. 
\end{itemize}

In a centralized topology, there is only one observability server. This server receives all the instrumentation data sent by the agents deployed on monitored devices or queries each node about the data of interest. 
This topology is more accessible to implement, making monitoring agents more straightforward, but it has disadvantages. Firstly, there is the \hl{Single Point of Failure} (SPOF) issue, where a server failure can interrupt the monitoring updates of the entire Fog environment and harm decision-making; secondly, the server must run in a resource-rich node to cope with data flow and storage, and its computational capacity will limit the number of monitored sources it is capable of handling \cite{zhao2017simmon}. A possible solution to this issue is to put the server in a cluster of Fog nodes, providing high availability. Lastly, server network channels can become overloaded with the observability data flow.

In a decentralized topology, there is at least one additional data staging layer between the monitored node and the observability server. The monitored sources are categorized by locality and assigned to a local observability server. This intermediary component may function as a centralized server for those nodes, where all observability data are transferred and stored longer than the device could.  In addition, it is possible to filter and aggregate the data before sending them to support decision making. However, this local server may not be the final destination of observability data. The observability server can be an independent component in the system, receiving and managing a high-level view of the observability data and integrating with a \hl{Fog service orchestrator} \cite{costa2021orchestration}, a Fog management component responsible for the Fog service life cycle that uses observability data to make decisions such as resource management and service offloading. Otherwise, the observability server can be implemented as a Peer-to-Peer (P2P) network of local servers, and the decentralized topology will function as a hybrid topology between the centralized and distributed ones.

In a distributed topology, the observability server is implemented as a P2P network of components distributed at all Fog nodes. The collected observability data must be shared (replicated) with all peers that need it and eventually with the Fog service orchestrator for resource and service management decision-making. 
Abderrahim \cite{abderrahim2017holistic} referred to this topology as the best option for Fog environments. It has the benefit of overcoming issues of observability server found on Centralized topology (f.i. SPOF, runs only on resource-rich nodes, network congestion). However, distributing this relevant function to some unreliable nodes can lead to partial and outdated data on device and service statuses. In addition, the more extensive the distributed server P2P network, the greater the risk of desynchronizing replicated observability data and the higher the \hl{communication overhead}. 

\section{Tools and Technologies for Observability on Fog Computing}
\label{sec:tools}
Observability in the IoT services scenario is related to the ability to perceive the environment by collecting instrumentation data and communicating those data as soon as possible for analysis and actuation. Several tools and technologies can contribute to achieving a high level of observability in this dynamic scenario, and the most common situation is a solution consisting of several different tools \cite{costa2022monitoring,karumuri2021towards}, increasing the complexity of management. In the following paragraphs, we present the most widely used open-source tools, focusing on the ones that can be deployed as components of a Fog observability system with an architecture like the one presented in Section \ref{sec:architecture}.

\subsection{Metrics Management}

Using metrics instrumentation in Fog Computing is essential for achieving comprehensive observability across distributed and heterogeneous environments. The metrics instrumentation domain has the highest number of tools available, probably because collecting hardware metrics using operating system calls is simpler, mainly in Linux distributions. Among these tools is Prometheus, which will be presented in the following.

\subsubsection{Prometheus Stack}
\hl{Prometheus}~\cite{prometheus} 
\index{Prometheus}%
is the de facto standard for collecting and storing metrics in the Cloud-native world. It created a data exposition format used by external agents, running on IoT devices or integrated into monitored systems, to generate metrics data and make them available for collection. Currently, dozens of agents can export metrics using this format, and several of them have characteristics suitable for use in Fog. NodeExporter\cite{prometheus} and cAdvisor \cite{noauthor_cadvisor_nodate} are examples of such metric exporters. The former exports mainly hardware and network metrics (CPU, memory, and disk usage, packets exchanged, among others) from the device. The later exports metrics of running containers. Prometheus can be configured to pull the metrics from those agents in a defined frequency, and all exposed metrics are scraped and stored in an internal TSDB. 

Prometheus data exposition format was used as a basis to create an industry-standard called \hl{OpenMetrics} \cite{openMetrics}.
\index{OpenMetrics}%
It is intended to be primarily a standardized format for the exposure of metrics, independent of any specific transport. The format is text-based, presenting each metric value in one line. It is easy to read, though it can be verbose when extracting many metrics. OpenMetrics is evolving to include enhancements and improvements to the Prometheus format, including additional metrics types and unification of escape rules. OpenMetrics provides programming language libraries to ease standard usage on applications and services, allowing easy creation of customized metric collection Agents.

\subsection{Logs Management}
In Fog Computing, log instrumentation is crucial in capturing and analyzing system events, errors, and operational data across distributed nodes. Several essential tools can be used to manage and analyze logs in these environments, including Filebeat and Fluentd, presented below.

\subsubsection{Filebeat}
The provided LaTeX text has been paraphrased as requested.
\hl{Filebeat} \cite{Filebeat} 
\index{Filebeat}%
operates as a local agent on IoT devices, monitoring changes in the file system, and transmitting new log data to an observability server. This intelligent agent allows users to designate specific files and folders of interest. Users can also set the frequency for checking each monitored file to efficiently manage the resource consumption on the device. Using a local cache improves data management and addresses temporary connectivity issues, resuming data flow once the connection is re-established.

\subsubsection{Fluentd}
\hl{Fluentd} \cite{Fluentd} 
\index{Fluentd}%
is an open-source platform that acts mainly as a log data management tool. Integrates all aspects of log data processing: collection, filtration, buffering, anomaly detection, and output routing across diverse sources and destinations. Its adaptable architecture is further enhanced by a wide array of plugins that broaden its functions, such as those that allow log data reception from agents and data exporting to storages such as ElasticSearch \cite{Elastic}, along with visualization tools such as Kibana \cite{kibana}. It can be identified as a Transformer when routing log data to an observability server or as a server component when serving as the final log data destination. Fluentd also offers robust failover mechanisms and can be configured to ensure high availability.

\subsubsection{Logtash}
\hl{Logtash} \cite{Logtash} 
\index{Logtash}%
is a tool akin to Fluentd in its capability to function as either a Transformer or a Server component in an observability system's architecture. It has numerous plugins and can interface with multiple agents for log data aggregation and storage solutions. The primary distinction between Fluentd and Logtash is that Logtash is less versatile due to its integration within the ELK stack, which includes ElasticSearch for data storage, Logtash itself, and Kibana. In contrast, Fluentd has a smaller memory and CPU footprint and excels in real-time data processing and event routing. Logtash is most effective in setups that already utilize the ELK stack.

\subsection{Traces Management}

In Fog Computing, trace instrumentation is carried out by several advanced tools designed to monitor and analyze the flow of requests across distributed systems. Jaeger and Zipkin are the most prominent tools for tracing in Fog environments. These tools are essential for maintaining the reliability and efficiency of Fog Computing infrastructures, where the seamless operation of distributed services is critical. These tools are described in the following.

\subsubsection{Jaeger}
\hl{Jaeger} \cite{Jaeger} 
\index{Jaeger}%
is an open-source trace management tool that helps to understand the flow of execution requests in services and applications. It is used to receive, store, and visualize traces and spans. It can be configured to use ElasticSearch as storage, among other database solutions, and provides proper visualization and fast query response time. 
Jaeger has no single point of failure and is scalable. It allows for the processing of several billion spans per day with modest computational power and is suitable for Fog environments. It has been designed from the ground up to support the OpenTracing standard and its instrumentation libraries.

\subsubsection{Zipkin}
\hl{Zipkin} \cite{Zipkin} 
\index{Zipkin}%
is a popular distributed tracing
system that is used to collect, look up, and visualize the tracing data of distributed systems. It has a simplistic, yet powerful user interface that enables the user to see the recent traces and execute queries against all the available traces. For example, traces can be filtered by service name,
duration, or timeframe. In addition, the Zipkin interface displays a graphical simulation of how requests logically flow from one service to another.

\subsection{OpenCensus, OpenTracing and OpenTelemetry}
Besides OpenMetrics, two other projects aimed at standardized telemetry data, but in this case more focused on traces: OpenCensus \cite{openCensus} and OpenTracing \cite{openTracing}. 
The two projects are vendor neutral, and even though their goals are somewhat similar, each follows a distinct path towards observability and instrumentation, leading to different evolutionary trajectories. Both OpenCensus and OpenTracing served as the basis for developing \hl{OpenTelemetry} (OTel) \cite{openTelemetry}, 
\index{OpenTelemetry}%
a standard designed to encompass the three pillars of observability.
OpenTelemetry is an open-source framework for observability that equips IT teams with standardized protocols and tools to collect and route telemetry data. Initiated as an incubator project by the Cloud Native Computing Foundation (CNCF), OpenTelemetry ensures a uniform format for instrumenting, generating, collecting, and exporting application telemetry data, including metrics, logs, and traces, to observability platforms for analysis.

OpenTelemetry comprises vendor-neutral open-source tools, APIs, and SDKs that can be implemented using various programming languages such as Go, Java, and Python. These tools collaborate to define what should be measured, gather the necessary data, clean and organize it, and then export it in suitable formats to an observability server. The components of OpenTelemetry are loosely coupled, allowing flexible integration of its different parts based on user requirements. However, because OTel is still a relatively new project, many features are not yet fully developed. For example, while distributed tracing is fully stable, many logging functionalities are still under development. Furthermore, OTel currently does not support automated analysis, as it lacks a sufficiently stringent specification \cite{bento2021automated}.


\subsection{Storage Tools}
Using data format standards and programming libraries such as OpenMetrics and OpenTelemetry facilitates the composition of solutions based on several independent elements. These elements can come from different vendors and be in different layers of the Fog Architecture. 
In addition to a standard data format for communication between different observability modules and solutions, those data can be stored on the IoT device for later transmission, stored on the Fog node for fast analysis and actuation, and stored in the Cloud for historical analysis, as described in steps 2, 4 and 6 of ODLC \ref{subsec:cycle} shown in Figure \ref{Fig:LifeCycle}. Table \ref{tab:domains} shows the differences of each instrumentation domain in the type of data and the most common queries made on them. Those characteristics require different storage types. Metrics use less space and are queried faster when stored in a time series database, such as InfluxDB \cite{influxdb} and QuestDB \cite{questdb}. Logs and traces are better supported by an inverted index database, such as \hl{ElasticSearch} \cite{Elastic}, 
\index{ElasticSearch}%
used to index, store, and quickly retrieve logs and traces data stored on it. To store and query high volumes of trace data, graph databases are also appropriate choices. ArangoDB and Neo4j are examples of such databases.

\begin{table}[h]
\centering
\addtolength{\tabcolsep}{-3.5pt}
\caption{Fog observability tools.}
\label{tab:software}
\begin{tabular}{@{}llcccl@{}}
\toprule
\textbf{Tool name} &
  \textbf{Domain} &
  \textbf{\begin{tabular}[c]{@{}c@{}}IoT \\ device\end{tabular}} &
  \textbf{\begin{tabular}[c]{@{}c@{}}Fog \\ node\end{tabular}} &
  \textbf{\begin{tabular}[c]{@{}c@{}}Cloud \\ server\end{tabular}} &
  \textbf{Observation} \\ \midrule
Node Exporter     & Metrics & \checkmark &      &      & \begin{tabular}[c]{@{}l@{}}Collects metrics from the device and exposes\\ them via an HTTP call.\end{tabular}               \\
cAdvisor     & Metrics & \checkmark &      &      & Collects metrics of running containers on a single machine.                \\
Filebeat          & Logs    & \checkmark &      &      & Log data collector.              \\
OpenTelemetry & Traces  & \checkmark &      &      & \begin{tabular}[l]{@{}l@{}}Programming language libraries to create and\\ transmit traces to observability servers.\end{tabular}    \\
Prometheus        & \hl{Metrics} &      & \checkmark & \checkmark & Pull metrics from the Node Exporter on each IoT Device. \\
InfluxDB        & Metrics &      & \checkmark & \checkmark & Time series database. \\
QuestDB        & Metrics &      & \checkmark & \checkmark & Time series database. \\
Fluentd          & \hl{Logs}    &  & \checkmark & \checkmark & Log management and processing.             \\
Fluentd          & Logs    &  & \checkmark & \checkmark & Log management and processing. Part of ELK stack.             \\
Elastic Search    & Logs,traces    &      & \checkmark & \checkmark & \begin{tabular}[l]{@{}l@{}}Inverted-index database that can stores logs \\and traces transmitted by collectors.\end{tabular}            \\
Neo4j        & Traces &      & \checkmark & \checkmark & Graph database. \\
ArangoDB        & Traces &      & \checkmark & \checkmark & Graph database. \\
Jaeger            & \hl{Traces}  &      & \checkmark & \checkmark & Receives and visualize trace data transmitted by collectors.                             \\
Zipkin            & Traces  &      & \checkmark & \checkmark & Receives and visualize trace data transmitted by collectors.                             \\
Kibana           &         &      & \checkmark & \checkmark & \begin{tabular}[l]{@{}l@{}}Presents dashboards of data from observability servers like \\Prometheus, Elastic Search and Jaeger.\end{tabular}     \\
Grafana           &         &      & \checkmark & \checkmark & \begin{tabular}[l]{@{}l@{}}Presents dashboards of data from observability servers like \\Prometheus, Elastic Search and Jaeger.\end{tabular}     \\ \bottomrule
\end{tabular}
\end{table}

\subsection{Visualization Tools}
To visualize the observability data, there are several open source dashboard solutions such as \hl{Grafana} \cite{Grafana}, 
\index{Grafana}%
Graphite \cite{Graphite} and Kibana \cite{kibana}, and specialized trace visualization such as Jaeger \cite{Jaeger}. These visualization tools are scalable, real-time charts rendering applications, written to visualize observability metrics gathered by other applications. They are capable of handling thousands of distinct metrics per minute \cite{kufel2016tools}. Their main purpose is to render graphs of stored observability data on demand. 

This Section listed some of the most referenced open-source observability tools. The tool name, the instrumentation domain that each tool supports, the Fog layer where they were deployed (IoT, Fog, or Cloud), and a brief description are consolidated in Table \ref{tab:software}.

\section{Solution Example}
\label{sec:solution}
A \hl{Smart City application}
\index{Smart City application}%
is designed to enhance and manage urban living standards. These applications utilize cutting-edge technologies such as IoT devices, 5G networks, and Artificial Intelligence to gather and evaluate real-time data on various aspects of the city, such as traffic, public safety, energy use, etc. ~\cite{gharaibeh2017smart}. Mobile IoT-RoadBot \cite{BANERJEE2024101326} is a Smart City application designed to monitor and identify issues related to roadside asset maintenance (dumped garbage and damaged bus shelters and road signs). This application has been implemented on 11 garbage collection trucks (GCT) operated by Brimbank City Council, Australia.  In addition, it serves as a practical testbed for evaluating the 5G performance on truck routes that cover more than 95\% of the 125 km$^2$ of the city area. It enabled the evaluation of 5G network performance through an application that transmits significant volumes of real-time video data over an extensive geographical region. The 11 GCTs have been equipped with IoT devices, including stereovision cameras, 5G dome antennas, fog nodes, and Global Navigation Satellite Systems.
Mobile IoT-RoadBot captures video along the roadside and streams data continuously while the GCT is operational for garbage collection across the city.

The application must compress video data at the edge and combine it with contextual information. The videos and supplementary data are then uploaded to the Cloud, where roadside assets are identified and integrated into the maintenance dashboard. GCTs employ 5G technology; however, 5G network coverage is not widespread. Consequently, there will be instances when a connection to the server is unavailable, necessitating the storage of videos for later transmission once the network becomes reachable. Furthermore, several issues, such as insufficient storage space for videos, camera failures, and application crashes, can disrupt the data flow between GCTs and cloud servers. Such delays in detecting these problems can result in data loss, financial waste, and reduced quality of service for citizens. The application writes a log line on the disk every second, recording some metrics collected from the sensors: truck speed, geographical coordinates, 5G throughput, and the amount of video data sent to the Cloud server, among others. As no observability system was deployed, these logs were analyzed asynchronously after the truck finished its daily routine.

\begin{figure*}[ht]
	\centering
	\includegraphics[width=\textwidth]{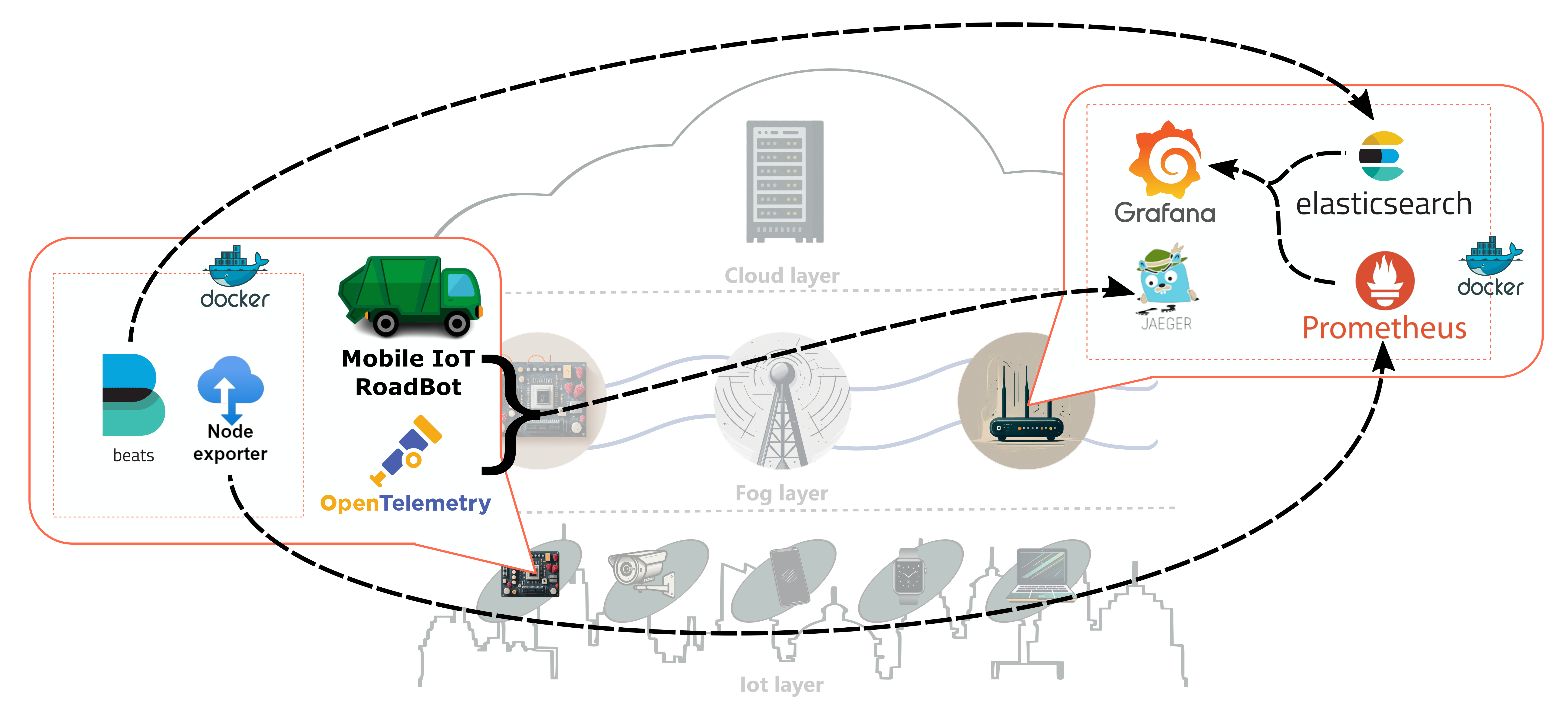}
	\caption{Fog observability system composed of open-source tools \cite{costa_achieving_2024}.}
	\label{Fig:Example}
\end{figure*}

Costa et al. \cite{costa_achieving_2024} deployed the \hl{Mobile IoT-RoadBot} \cite{BANERJEE2024101326} application on a \hl{Fog testbed} to evaluate the potential benefits and setbacks of using an observability system in this environment. The testbed comprises 4 IoT devices, two Fog nodes, and one virtual machine in the Cloud. The authors selected open-source solutions to collect metrics, logs, and traces from IoT devices and transmit them to a Fog observability server where data could be evaluated and generated in a timely manner. NodeExporter, Filebeat, and OpenTelemetry SDK were deployed in the IoT devices to collect metrics, logs, and traces, respectively. In the Fog nodes (and in the virtual machine on the Cloud), Prometheus, ElasticSearch, Jaeger, and Grafana were deployed to receive observability data, store them, and allow data cross-analysis and visualization.
Every tool was deployed using \hl{Docker containers}. All tools used were already described in Section \ref{sec:tools} and are listed in Table \ref{tab:software}. The devices hosting each piece of software and their observability data flow can be seen in Figure \ref{Fig:Example}. The virtual machine on the Cloud mimics the environment of the Fog node and is not shown in Figure \ref{Fig:Example} for simplicity.

The experiments measured the overhead that the observability tools added to the IoT devices and Fog nodes, as well as the volume of observability data collected and transmitted from the IoT Layer to the Cloud Layer.
They observed low CPU (less than 12\%) and memory (less than 150 MiB) overhead in the IoT devices needed to run all three components, that is, the NodeExporter, Filebeat, and OpenTelemetry SDK. 
However, in the Fog node, the CPU overhead reached 25\% to run the server-side components, i.e. Prometheus, Elastic Search, Jaeger, and Grafana. The memory footprint on the Fog node was 5GiB, requiring more resourceful devices to support them. Regarding observability data volume, the authors manually limited the metric set collected from the IoT devices to be only the ones of interest to the application, i.e., metrics related to CPU, memory, disk and network usage, and available power supply. Also, they restricted the maximum time for observability data to remain on the Fog node for one week.

After one week of age, data are transferred from the Fog node to the Cloud for long-term storage and historical analysis. These strategies resulted in 2GB of aggregated volume of observability data in a week. This volume would represent less than 1\% of the amount of data sent to the Cloud using the 5G network in a week of real-world operation. The experiments showed that it is possible to collect the benefits of achieving a higher level of observability for a system in a Fog computing environment with a low overhead in terms of resource usage.

\section{Future Directions}
\label{sec:future}
Observability is not a new term or subject \cite{kalman1960general}, but its meaning has evolved in the context of IoT services supported by Fog environments \cite{costa_achieving_2024,karumuri2021cloud,sigelman2010dapper}. It is a hot topic, and several works are trying to figure out how to use it to get the highest value with the lowest cost, not only in financial terms but also in terms of resource usage. This section presents some future directions for observability systems running in Fog environments.

The resources used by an observability system compete with those needed by the domain application and services \cite{costa2022monitoring}. This means that an observability system for \hl{Fog environments} should maintain a low footprint of computational resource consumption since IoT devices and \hl{Fog nodes} can be resource-restricted. A strategy to reduce resource usage in the ODLC Collection phase is to use the enhanced Berkeley Packet Filter (\hl{eBPF}) \cite{gregg2019bpf}. eBPF allows the execution of custom programs in a secure virtual machine in the operating system's kernel of the IoT device, providing a vantage point to monitor system-level activities. eBPF can be a very efficient tool for monitoring the system's behavior and detecting failures \cite{levin2020viperprobe}. Tracing is an application-specific effort, and deploying a single integrated tracing infrastructure across multiple application platforms and middleware components is challenging \cite{neves2021detailed}. Rezvani et al. \cite{rezvani2024characterizing} have explored using eBPF to provide not only hardware and network metrics but also application-level metrics for user-facing and latency-sensitive workloads. However, although eBPF is mature in the Linux scenario, it is not yet at the same level of evolution as other production platforms like Windows servers.

Interchangeable data formats may allow the connection of several observability data collection tools, such as collected, statsd \cite{noauthor_statsd_nodate}, sensu \cite{porter2016sensu}, Amazon CloudWatch, with commercial or open-source observability servers, which can accumulate data from several distinct sources, providing a more comprehensive view of the environment for maintenance and operational staff \cite{ward2014observing}.

The utilization of Artificial Intelligence is another future direction that can have a relevant impact on the Fog observability field. Large volumes of data can hinder users' ability to pinpoint system performance bottlenecks and impact overall system performance. Thus, selecting only the most relevant metrics is essential to reduce observability overhead, reduce costs, and improve performance analysis \cite{popiolek2021low}. However, identifying these key metrics usually relies on professional expertise. Furthermore, the significance of specific performance metrics might fluctuate over time with varying application workloads, rendering manual selection of the most relevant counters impractical. Popiolek et al. \cite{popiolek2021low} have proposed an approach that uses linear correlation and hierarchical clustering analysis to reduce the dimensionality of the data through feature selection.  A smaller number of higher-quality metrics can be more valuable than collecting a slew of data just because it is possible. Another item where Artificial Intelligence can add value is the support of adaptive behaviors \cite{bridges_gartner_2024}, replacing the primary threshold-based actuation for adaptive anomaly detection. An anomaly detection-based notification strategy supports the establishment of a baseline of utilization over time. For example, a two-hour spike in 90\% CPU usage on the last day of the month can be a regular event of the month-end processing and could not generate an alert like when it happens at the beginning of the month.

\section{Summary and Conclusions}
\label{sec:summary}
This chapter presented the relevant concepts and characteristics of observability systems running in a Fog Computing environment. A higher observability of a service or application means that more information about its internal execution is available. When analyzed in a timely manner, this information allows one to understand the root cause of issues in the runtime environment and helps solve them more quickly, increasing service availability. 

Fog environments are composed of \hl{resource-restricted devices} connected by potentially \hl{unreliable communication channels}. Thus, an additional load can be harmful if not properly managed. Then, a \hl{Fog Observability data lifecycle} is presented. It was proposed to guarantee that the Fog environment will not be overloaded with the observability data flow. The Lifecycle encompasses data collection from IoT devices, limited storage and fast analysis on Fog nodes, and the transmission to the Cloud of data outside a defined time window of interest for the Fog.

The data will be used in the Cloud for historical analysis and long-term storage.
The architecture  of observability systems on Fog was presented concerning the structural functions they need to implement as well as the different topologies they can be deployed.
We presented tools and technologies to help build robust and efficient observability systems for Fog Computing. These tools address challenges related to data format standardization, data storage and visualization, and analysis techniques. These tools and technologies empower organizations to gain meaningful insights from their Fog Computing systems while ensuring performance, reliability, and seamless user experiences. 
An example of a real-world solution was described to help the audience understand the concepts presented. Finally, future directions are discussed to broaden the vision for Observability in Fog Computing.

\section*{Acknowledgements}
This study was partially financed by the Coordenação de Aperfeiçoamento de Pessoal de Nível Superior (CAPES) - Brasil - Finance Code 001. Gratitude also goes to the BioCloud2 project, approved under CNPq/AWS Call No. 064/2022, CNPq process No.421828/2022-6.

\bibliography{chapter}

\end{document}